\documentclass[useAMS,usenatbib]{mn2e}
\usepackage{amsmath}
\usepackage{graphicx}
\usepackage{txfonts}

\title[Asymmetric structure in Sgr\,A* at 3\,mm from closure phase measurements with VLBA, GBT, and LMT]{Asymmetric structure in Sgr\,A* at 3mm from closure phase measurements with VLBA, GBT and LMT}
\author[C.D.Brinkerink et
al.]{\parbox{\textwidth}
{Christiaan~D.~Brinkerink$^{1}$\thanks{E-mail: \texttt{c.brinkerink@astro.ru.nl}},
Cornelia~M\"uller$^{1,2}$,
Heino Falcke$^{1}$,
Geoffrey C. Bower$^{3}$,
Thomas P. Krichbaum$^{2}$,
Edgar Castillo$^{7,8}$,
Adam T. Deller$^{4}$,
Sheperd S. Doeleman$^{5,6}$,
Raquel Fraga-Encinas$^{1}$,
Ciriaco Goddi$^{1}$,
Antonio Hern\'andez-G\'omez$^{9}$,
David H. Hughes$^{8}$,
Michael Kramer$^{2}$,
Jonathan L\'eon-Tavares$^{8,10}$,
Laurent Loinard$^{2,9}$,
Alfredo Monta\~na$^{7,8}$,
Monika Mo\'scibrodzka$^{1}$,
Gisela N. Ortiz-Le\'on$^{9}$,
David Sanchez-Arguelles$^{7}$,
Remo P. J. Tilanus$^{1,11}$,
Grant W. Wilson$^{12}$,
J. Anton Zensus$^{2}$}
\\
\parbox{\textwidth}{\vspace{0.4cm}
$^{1}$Department of Astronomy, IMAPP, Radboud University, Heyendaalseweg 135, 6525 AJ Nijmegen, the Netherlands 
$^2$Max-Planck-Institut f{\"u}r Radioastronomie, Auf dem H{\"u}gel 69, 53121 Bonn, Germany 
$^3$Academia Sinica Institute of Astronomy and Astrophysics, 645 N. A'ohoku Pl., Hilo, HI 96720, USA 
$^4$ASTRON, the Netherlands Institute for Radio Astronomy, Postbus 2, 7990 AA, Dwingeloo, The Netherlands  
$^5$MIT Haystack Observatory, Off Route 40, Westford, MA 01886, USA  
$^6$Harvard Smithsonian Center for Astrophysics, 60 Garden Street, Cambridge, MA 02138, USA
$^7$Consejo Nacional de Ciencia y Tecnolog\'ia, Av. Insurgentes Sur 1582, Col. Cr\'edito Constructor, Del. Benito Ju\'arez, C.P.: 03940, D.F., M\'exico
$^8$Instituto Nacional de Astrof\'isica \'Optica y Electr\'onica (INAOE), Apartado Postal 51 y 216, 72000, Puebla, M\'exico
$^9$Instituto de Radioastronom\'ia y Astrof\'isica, Universidad Nacional Aut\'onoma de M\'exico, Morelia 58089, M\'exico
$^{10}$Sterrenkundig Observatorium, Universiteit Gent, Krijgslaan 281-S9, B-9000 Gent, Belgium
$^{11}$Leiden Observatory, Leiden University, P.O. Box 9513, 2300 RA Leiden, The Netherlands
$^{12}$University of Massachusetts, Department of Astronomy, LGRT-B 619E, 710 North Pleasant Street, Amherst, MA 01003-9305, USA
}}
\begin{document}

\date{\today}

\pagerange{\pageref{firstpage}--\pageref{lastpage}} \pubyear{2015}

\maketitle

\label{firstpage}

\begin{abstract}
We present the results of a closure phase analysis of 3\,mm very long baseline interferometry (VLBI) measurements performed on Sagittarius\,A* (Sgr\,A*). We have analyzed observations made in May 2015 using the Very Long Baseline Array, the Robert C. Byrd Green Bank Telescope and the Large Millimeter Telescope Alfonso Serrano and obtained non-zero closure phase measurements on several station triangles - indicative of a non-point-symmetric source structure. The data are fitted with an asymmetric source structure model in Sgr\,A*, represented by a simple two-component model, which favours a fainter component due East of the main source. This result is discussed in light of a scattering screen with substructure or an intrinsically asymmetric source.
\end{abstract}


\section{Introduction}

The supermassive black hole candidate at the center of our Galaxy (associated with the radio source Sagittarius\,A*, or Sgr\,A*) offers a prime possibility to study the physical phenomena associated with accretion onto a supermassive black hole \citep{Genzel2010, Falcke2013, Goddi2016}. Sgr\,A* is thought to accrete at an extremely low Eddington ratio \citep{Falcke93, Quataert00}, an accretion regime analogous to the low-hard state in X-ray binaries and for which a jet component is expected to manifest. These expected physical behaviours and their interplay make it challenging to formulate fully self-consistent models for Sgr\,A* that simultaneously explain its spectrum, its variability and its size and shape on the sky. The expected angular size of the event horizon of Sgr\,A* on the sky \citep[50 $\mu$as,][]{Falcke2000} is the largest of any known black hole candidate. This makes it a prime target for studies using very long baseline interferometry at mm wavelengths (mm-VLBI), which can attain spatial resolutions that are comparable to the expected shadow size on the sky \citep{Doeleman2008, Falcke2013}.\\

\noindent
A second reason to use VLBI measurements at short wavelengths is due to the interstellar scattering that is encountered when looking at the Galactic Center in radio \citep{Backer1978}. Sgr\,A* exhibits an apparent size on the sky that is frequency-dependent, scaling with $\lambda^2$ \citep[the exact exponent depends on the specific type of turbulence in the interstellar plasma, see][]{Lu2011} for observing wavelengths longer than about 7\,mm \citep[43\,GHz, ][]{Bower2006}. This is due to interstellar scattering by free electrons: at these wavelengths, the scattering size is significantly greater than the intrinsic source size and as such the apparent source size is dominated by the scattering effect. At wavelengths shorter than 7\,mm, the apparent source size breaks away from from the $\lambda^2$-relation and the intrinsic source size can be more easily recovered after quadrature subtraction of the known scattering size for that wavelength \citep{Falcke2009}. The shorter the observing wavelength, the more prominent the intrinsic source size and shape shine through. The relation between the intrinsic source size (i.e., the size after correcting for the scattering effect) and the observing wavelength has also been investigated, showing that the emission region itself shrinks with decreasing observing wavelength too. At an observing wavelength of 1.3\,mm (230\,GHz), the size of Sgr\,A* on the sky has been shown to be even smaller than the expected projected horizon diameter of the black hole \citep{Doeleman2008}.\\

\noindent
The present view is that the cm- to mm-wavelength spectrum of Sgr\,A* is generated by partially self-absorbed synchrotron emission from hot plasma moving in strong magnetic fields close to the putative event horizon of the black hole, a model supported by recent observations and analyses thereof \citep[][ and references therein]{Doeleman2008, Fish2011, Lu2011, Bower2014, Gwinn2014, Fish2016, Broderick2016}. See \citet{Falcke2013} for a recent review on our current understanding of the nature of Sgr\,A*. However, the specific part of the black hole environment where this emission is thought to come from is subject to debate. Many properties of the bulk accretion flow such as density, temperature and magnetic field strength can be investigated using general-relativistic magnetohydrodynamic (GRMHD) simulations, and results from different modern simulations paint a consistent picture. However, much depends on the specific prescription for the electron temperature that is used throughout the accretion flow. For Sgr\,A*, the inner region of the accretion disk has been put forward as the main emission region candidate if certain electron temperature prescriptions are used \citep[e.g.,][]{Narayan1995}, but other physically motivated prescriptions indicate that the jet launching region may dominate mm-wavelength emission instead \citep[e.g.,][]{Moscibrodzka2013}. These different models yield comparable predictions for the expected overall size of the source at 86\,GHz, but predict different source shapes.\\

\noindent
To resolve this debate, gathering more accurate knowledge of the detailed brightness distribution of the source on the sky (particularly its asymmetry) plays an important role. Observations at 3.5\,mm (86\,GHz) provide an excellent way of studying this geometry: the emission comes from the inner accretion region, but it is not so strongly lensed as the 1.3\,mm emission is thought to be. This means that the apparent source shape at 3.5\,mm provides the best insight into which regions of the inner accretion flow form the source of the radiation that we receive.\\

\begin{figure}
\includegraphics[width=0.48\textwidth]{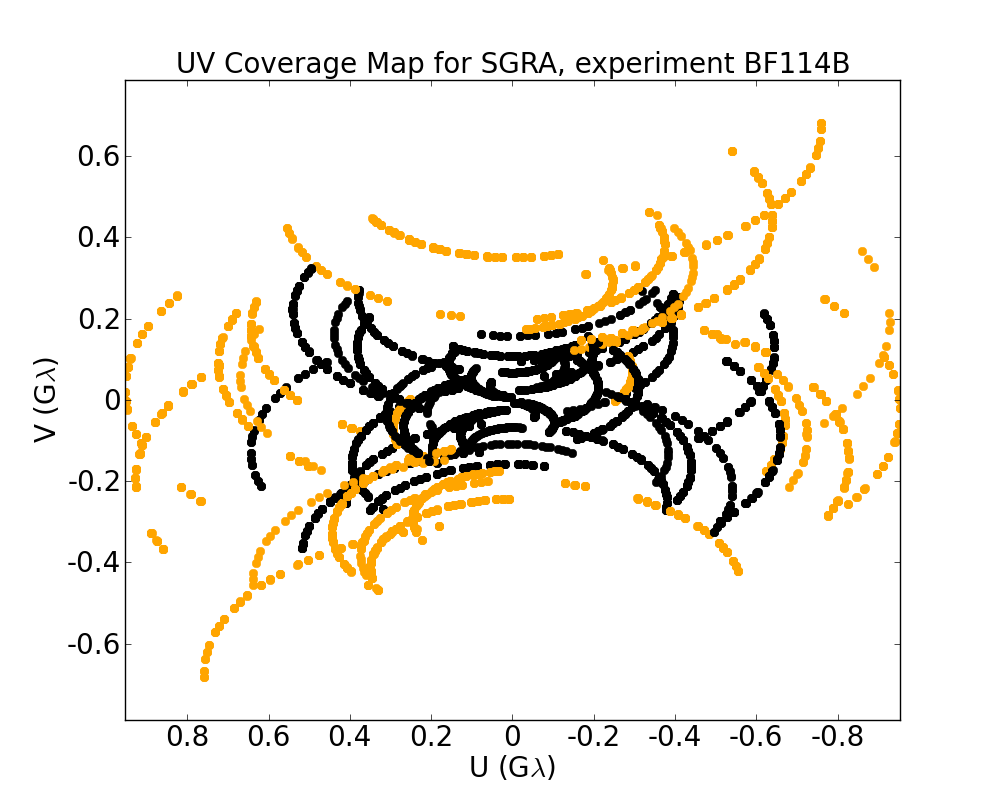}
\caption{The $(u,v)$ coverage for the observation of Sgr\,A* taken on May 23rd, 2015 (6:00 - 13:00 UT). Baselines within the VLBA are coloured black, baselines to LMT and GBT are coloured orange. No baselines to Mauna Kea (MK) are shown, as we have not found fringes for Sgr\,A* on any baseline to MK. The inclusion of LMT improves North-South $(u,v)$ coverage, while the inclusion of GBT improves East-West coverage.}
\label{uvcoverage}
\end{figure}

\noindent
The only telescope arrays that can reach the angular resolution on the sky required to potentially discern this asymmetry are the High-Sensitivity Array (HSA), the Global mm-VLBI Array (GMVA), and the Event Horizon Telescope \citep[EHT,][]{Doeleman2008}.  Before 2015 these VLBI arrays offered limited North-South $(u,v)$ coverage for Sgr\,A*, which is in the Southern sky (RA: 17h45m40s, DEC: -29d00m28s), and have thusfar left the question of asymmetric source structure open. With the inclusion of the Large Millimeter Telescope Alfonso Serrano (LMT) in the HSA as of the first semester of 2015 \citep[see][for the description of VLBI implementation at LMT]{Ortiz2016}, the $(u,v)$ coverage at 3.5\,mm has been improved dramatically (see Fig.~\ref{uvcoverage}).\\

\noindent
Using observations at longer wavelengths (ranging from 7\,mm to 6\,cm), for which interstellar scattering dominates the observed source size, it has been shown that the scattered source has an elongated, approximately Gaussian structure \citep{Shen2005, Bower2006-2} with major and minor axes that scale with observing wavelength as $b_{\textrm{maj,scatt}} = 1.32 \pm 0.02$ mas cm$^{-2}$ and $b_{\textrm{min,scatt}} = 0.67 \pm 0.02$ mas cm$^{-2}$ respectively \citep{Bower2015}. This observed Gaussian has a well-defined position angle of $81.8^{\circ}\pm 0.2^{\circ}$ East of North. Extrapolated to $\lambda$=3.48\,mm, this relation yields a scattering size of $(160\pm2) \times (81\pm2)\,\mu$as. Recent measurements at 3.5\,mm, done with the VLBA and the LMT, indicate that the observed size is $(216\pm5) \times (143\pm8)\,\mu\rm{as}$, at a position angle of $80.5^\circ\pm4^\circ$ East of North - indicating that the intrinsic structure of Sgr\,A* is partially resolved and yielding an estimate for the intrinsic size after quadrature subtraction of the scattering size of $(147\pm7) \times (120\pm12)\,\mu$as at a position angle $80^\circ \pm 7^\circ$ \citep[][note that we quote the more conservative closure amplitude derived results here]{Ortiz2016}. Moreover, the closure phases measured by that work are mentioned to be consistent with the expected values produced from the effects of interstellar scattering alone, although the cause for the non-zero closure phases may yet be intrinsic to the source.

\noindent
Some recent results do suggest the presence of (possibly time-variable) asymmetry in Sgr\,A*, however. Persistent source asymmetry for Sgr\,A* has been measured at 230\,GHz in observations by the EHT, where an East-West asymmetry is suggested by simple model fitting results \citep{Fish2016}. Tentative evidence for (transient) source asymmetry has also been seen in observations from 2012 at 43\,GHz, as reported by \citet{Rauch2016}, where one 2-hour subinterval in an 8-hour observation showed a secondary South-Eastern source component at a separation of approximately 1.5\,mas. This timescale is too short for the perceived structural variation to be due to changes in the scattering screen, and would point to intrinsic structural change in the source. However, the significance of this secondary component is quoted to be at the 2-$\sigma$ level.\\

\noindent
In this work, we present our first findings obtained from observations of Sgr\,A* at 3\,mm, involving the Very Long Baseline Array (VLBA), the Green Bank Telescope (GBT) and the newly added Large Millimeter Telescope (LMT) in Mexico. Section\,2 details the observations, as well as the data reduction steps performed. In Section\,3, we discuss possible instrumental causes for non-zero closure phases and verify that our observations are not significantly affected by them.  Section\,4 presents the measured closure phases and the model fit results. Section\,5 contains discussion on the results and offers our interpretation of them. Finally, our conclusion is stated in Section\,6.

\section{Observations and initial data reduction}

We present our analysis based on data from a single epoch of 3\,mm HSA observations, which was recorded on May 23rd, 2015 (5:00 to 14:00 UT, project code BF114B). The track has the VLBA together with LMT and GBT as participating facilities. Of the VLBA, the following stations were involved in the observation: Brewster (BR), Fort Davis (FD), Kitt Peak (KP), Los Alamos (LA), Mauna Kea (MK), North Liberty (NL), OVRO (OV) and Pie Town (PT). Only left-circular polarization data was recorded, at a center frequency of 86.068\,GHz and a sample rate of 1024\,Ms/s (2-bit) - this translates to an effective on-sky bandwidth of 480\,MHz, which is divided up into 16 IFs of 32\,MHz each. The 16th IF falls partly out of the recording band and was flagged throughout our dataset. We used 3C\,279 and 3C\,454.3 as fringe finder sources. Our check-source and secondary fringe finder was NRAO\,530, and observations were done in scans of 5 minutes, alternating between NRAO\,530 and Sgr\,A* for most of the tracks. Pointing for the VLBA was done at 43 GHz on suitable SiO masers every half hour, while the LMT and GBT did their pointing independently during the same time intervals (taking $\sim$10 minutes). For the VLBA pointing solutions, we assumed that the offset between the optical axis at 3\,mm and at 7\,mm for each station antenna had remained stable since the last calibration run done before our observation.\\

\noindent
The data were correlated with the VLBA DiFX software correlator (v. 2.3) in Socorro, and initial data calibration was done in AIPS \citep{Greisen2003}. System temperature ($\mathrm{T_{sys}}$) measurements and gaincurves for LMT and GBT were imported separately, as they were not included in the a-priori calibration information provided by the correlator. Edge channels in each IF were flagged (five channels on each side out of 64 channels, corresponding to $\sim$16\% of the subband), and the AIPS task \texttt{APCAL} was used to solve for the receiver temperatures and atmospheric opacity and to set the amplitude scale. In the initial \texttt{FRING} step, we used the primary fringe finder scans to correct for correlator model delay offsets and for the delay differences between IFs ('manual phasecal'), the solutions of which were then applied to all scans in the data set. The second \texttt{FRING} run solved for the delays and rates for all sources, using a solution interval of two minutes, while combining all IFs (\texttt{APARM(5) = 1}). Failed solutions that were flagged by the \texttt{FRING} task (about 10\% of the total) were left out for the remainder of data reduction. No fringes on baselines to MK were found, but all other baselines did yield clear detections. At this point, the fringe-fitted data were fully frequency-averaged (channel-averaged and IF-averaged) to a single channel, and exported to UVFITS and loaded into Difmap \citep{Shepherd1997}.\\

\noindent
Low source elevations during the observation can in principle cause the atmospheric coherence time to be very limited, leading to a loss of signal quality when time-averaging data that has been calibrated too coarsely in time. To verify that coherence issues would not be affecting our data quality, separate \texttt{FRING} runs were done with solution intervals shorter than two minutes. The length of the solution intervals in this range was found to have no significant impact on the later derived closure phase values, only increasing their uncertainties. Shorter solution intervals for FRING resulted in a larger fraction of failed solutions.\\

\noindent
Without an accurate a-priori model of a source, phase and amplitude calibration in VLBI is notoriously tricky: the amplitude uncertainties after calibration can be as large as 10\%-30\% for VLBI data at 3\,mm wavelengths \citep{Martividal2012}. The main reason for this is incomplete knowledge of the gain-elevation dependences, the presence of residual antenna pointing and focus errors and the highly-variable atmosphere, for which the applied opacity correction only partially corrects the time-variable absorption. For this reason, our primary goal was to look at quantities which are not station-based and which are free from local gain variations. The closure phase is such a quantity.\\

\noindent
Closure phase is the phase of the product of visibilities (equivalently, the sum of phases) taken from three connected baselines forming a triangle where station order is respected \citep{Jennison1958}. Closure phases are unaffected by station-based phase fluctuations, which are typically caused by tropospheric delays due to variable weather, clock drifts from the local maser, or time-dependent characteristics of the receiver system. Such station-based phase offsets cancel out when forming the closure phase. See \citet{Rogers1995} for an extended discussion on the characteristics of closure phase uncertainties.\\

\noindent
We used Difmap to time-average the fringe-fitted, frequency-averaged data as exported from AIPS from 0.5-second integrations into 10-second blocks (command: \texttt{uvaver 10, true}). This step was also tested with different averaging intervals, and the 10-second interval was found to yield the highest signal-to-noise ratio (SNR) for the eventual closure phase measurements. The time averaging was done to obtain a higher SNR per datapoint, while respecting the coherence time of the atmosphere ($\sim$10\,s to 20\,s at 86\,GHz). Longer time averaging intervals (15s, 30s) were found to yield compatible results, but with slightly worse noise characteristics. We chose not to phase-selfcalibrate the data in AIPS (beyond fringe-fitting at the two-minute timescale) before this step, for the main reason that it would result in a significant fraction (over 50\%) of the remaining visibilities being flagged because of failed solutions from low SNR. Instead, we chose to use the closure phases derived from the 10-second averaged data directly in the subsequent stage of data reduction. The use of closure phases sidesteps the (station-based) noise issues associated with individual visibilities, avoiding a large source of error in the resulting data. The second rationale for this approach is that we wanted to perform this analysis in as much a model-independent way as possible. To assess the possible influence of frequency-dependent data artefacts on the calculated closure phases, the closure phase calculations were also done using exported data from AIPS where all 15 IFs were kept separate and in which each IF was channel-averaged. This alternative method was found to yield fully compatible closure phase values, but with slightly larger closure phase errors.\\

\noindent
The SNR for each time- and IF-averaged closure phase measurement was high enough to avoid the potential issue of phase wrapping when averaging. We therefore averaged the closure phase measurements using error-weighted summation on the phase values. We estimated the associated error on the averaged value according to $\sigma_{\textrm{cp}} = \sigma_{\textrm{scan}} / \sqrt{n}$, where $\sigma_{\textrm{scan}}$ is the standard deviation of the observed closure phase distribution over one scan and $\sqrt{n}$ is the square root of the number of measurements averaged within one scan (typically, $n\approx30$). If the SNR per measurement were too low, the occurrence of phase wrapping when averaging the closure phase values would bias the result towards zero.\\

\begin{figure*}
\includegraphics[width=0.48\textwidth]{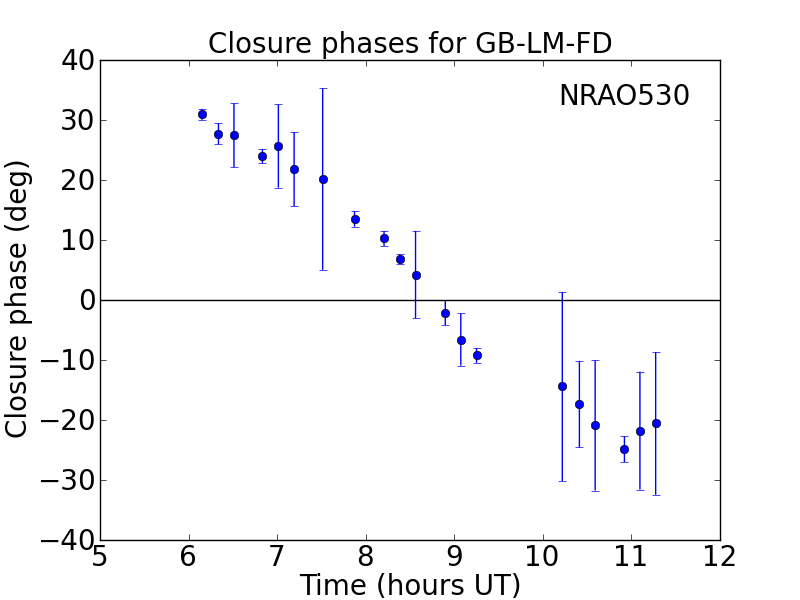}
\includegraphics[width=0.48\textwidth]{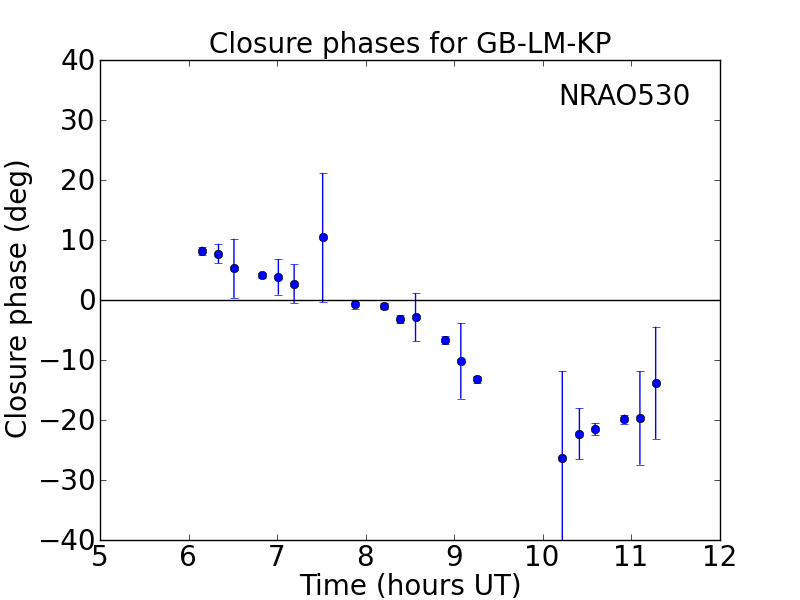}
\includegraphics[width=0.48\textwidth]{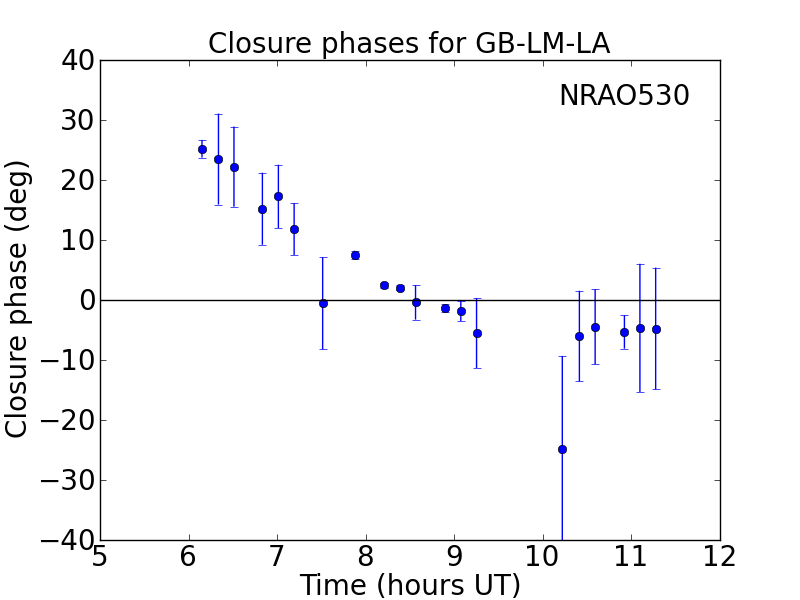}
\includegraphics[width=0.48\textwidth]{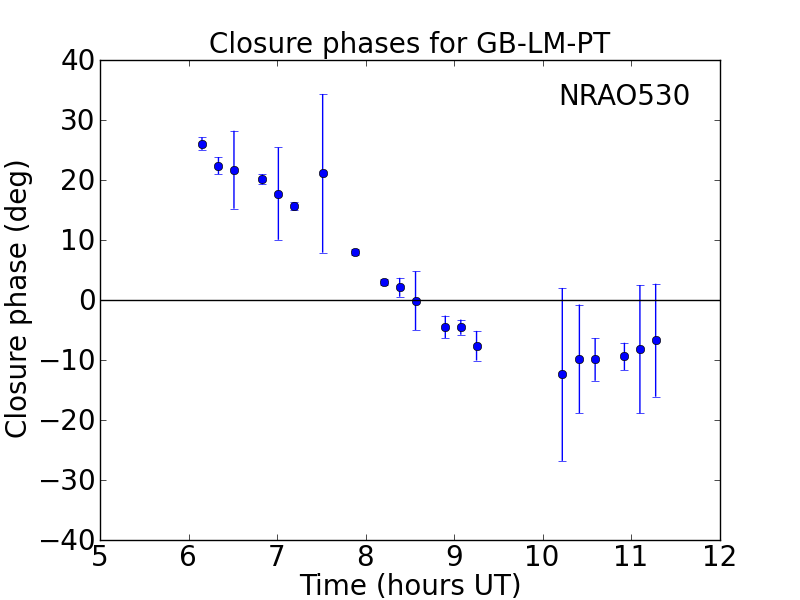}
\caption{Closure phase versus time for NRAO\,530 for the four `central' triangles in the array. The closure phase depends nonlinearly on the East-West extent of the triangle, which is the reason the GB-LM-KP triangle exhibits a somewhat different closure phase evolution.}
\label{closurephases-nrao530}
\end{figure*}

\noindent
NRAO\,530 exhibits known asymmetry in source structure \citep{Bower1997,Lu2011}. Using NRAO\,530 as a check-source for our closure phase measurements, we recover clear closure phase trends over time on most station triangles (see Fig.~\ref{closurephases-nrao530} for the clearest of these). When we apply the same averaging scheme to the closure phase measurements for Sgr\,A*, we see that the clearest closure phase measurements - with the highest SNR - are typically obtained on triangles that have both LMT and GBT as participating stations (see Fig.\,\ref{closurephases-sgra}).  All of these triangles show closure phase deviations away from zero with consistent sign, suggesting an asymmetry in the source image for Sgr\,A*. The relatively large closure phases measured for the GBT-LMT-KP triangle in comparison with the other triangles shown is a natural consequence of the greater East-West extent of this triangle: the model-fitting results (discussed in Section 4) show the same relatively large closure phases on this triangle, as indicated by the continuous lines in the plots. We have verified that this larger closure phase variation is not due to the station performance at KP by studying the stability of the amplitudes and phases of the visibilities on baselines to KP obtained close to the 7:00 - 8:00 UT time interval, the fringe fitting solutions (delay and rate), the bandpass response, as well as the atmospherical stability and system temperature behaviour. None of these parameters showed aberrant behaviour.\\

\noindent
Measurements with high SNR are also obtained on small and `degenerate' triangles. Degenerate triangles are triangles that have one short baseline on which the fringe spacing on the sky is much larger than the scattered source size, and for which the visibility has an expected phase of zero. This high SNR is expected due to the large visibility amplitudes that these triangles have on their short baselines. We find that the triangles involving VLBA stations NL or OV show the lowest SNR. In the case of NL triangles, this is likely caused by the low maximum elevation of Sgr\,A* in the local sky. For OV it is likely due to the bad weather causing high (and rapidly fluctuating) atmospheric opacity at the site on the day of the observation.\\

\section{Verifying the nature of non-zero closure phases}

There is a danger that non-zero closure phases can be caused by various instrumental causes. Phase variations in the bandpass can potentially cause non-zero closure phases for a point source. This closure phase bias is given by:
\begin{equation}
\begin{split}
\Delta \phi_{\textrm{BP}} = \textrm{arg}\left( \int g_1(\nu)g_2^*(\nu) d\nu \right) + \textrm{arg}\left( \int g_2(\nu)g_3^*(\nu) d\nu \right) + \\
\textrm{arg}\left( \int g_3(\nu)g_1^*(\nu) d\nu \right),
\end{split}
\end{equation}

\noindent
where the $g_{i}(\nu)$-terms are the complex frequency-dependent gains for antenna $i$, and the integral is performed over the full observed frequency band. We checked for phase slopes across all IFs by running the AIPS task \texttt{BPASS} on our check-source NRAO\,530 (to obtain a high SNR) with a solution interval of 5 hours, and the resulting bandpass correction does not exhibit phase slopes of more than 20\,degrees across the full 0.5\,GHz bandwidth for any station. By simulating point source data observed by one triangle and introducing a range of thermal noise and different phase slopes across the band for one antenna, we have separately verified that phase slopes below 2\,radians ($\sim$116\,degrees) over the full bandwidth have no significant effect on the measured closure phases. Closure phase measurements taken in separate subbands also show results highly consistent with what we see from combined subbands. We are therefore confident that the closure phases we see are not caused by bandpass calibration irregularities.\\

\noindent
Another possible instrumental cause for non-zero closure phases is the presence of polarization leakage for significantly linearly polarized sources.  Although our observation is LCP only, the RCP component of incoming radiation bleeds into the LCP signal chain in a limited way, and this may cause anomalous closure phases. The expression describing the closure phase bias from polarization leakage is given by:
\begin{equation}
\begin{split}
\Delta \phi_\textrm{pol} = \textrm{arg} \left( 1 + P \cdot D_1 \exp{i \psi_1} + (P \cdot D_2 \exp{i \psi_2})^* \right) + \\
\textrm{arg} \left( 1 + P \cdot D_2 \exp{i \psi_2} + (P \cdot D_3 \exp{i \psi_3})^* \right) + \\
\textrm{arg} \left( 1 + P \cdot D_3 \exp{i \psi_3} + (P \cdot D_1 \exp{i \psi_1})^* \right)
\end{split}
\end{equation}
\noindent
In the above expression, $P$ represents the linear polarization fraction of the source while $D_i$ is the complex polarization leakage term from RCP to LCP for a given antenna i. $\psi_i$ is the difference between the position angle on the sky of the source polarization vector and the parallactic angle of antenna $i$. When we use an upper bound for the correlated linear polarization of Sgr\,A* at 3\,mm as being 2\% \citep{Bower1999, Macquart2006} and the magnitude of the complex D-terms as being at most 10\% \citep[as indicated by recent GMVA results: see Table 1 in][]{Martividal2012}, we get negligible leakage-induced closure phase errors ($\Delta \phi_\textrm{pol} < 6 \cdot 10^{-4}$ deg) if we let the $\psi_i$-values vary so as to get the maximum possible polarization leakage.

\section{Results}

\subsection{Detection of non-zero closure phases}

The four triangles formed by the LMT, GBT and one of the four southwest VLBA stations (FD, KP, LA or PT) are the triangles that show the clearest evolution of closure phase with time. They suggest closure phase trends for Sgr\,A* with time that seem mutually compatible (see Fig.~\ref{closurephases-sgra}), due to the roughly similar orientations and lengths in the $(u,v)$ plane probed by their baselines. We note that the magnitude of the closure phase deviation from zero depends on the extent of the triangles in a nonlinear fashion, as was tested for a range of triangle geometries using simple source models with asymmetry, potentially explaining why the closure phases on the GBT-LMT-KP triangle are larger than those seen on the other triangles in the plot. However, all of these triangles show closure phase deviations away from zero in the same direction.\\

\subsection{Modeling source asymmetry using closure phases}
\noindent
We wish to investigate the possible presence of point asymmetry of Sgr\,A* using closure phase measurements. The simplest model that can exhibit any asymmetry and non-zero closure phases is a model using two point source components, with the two components having unequal flux densities. Although the average scattering ellipse would suggest that any point-like source components should show up as 2D Gaussians, the fact that the scattering screen itself can impose substructure on even smaller angular scales provides additional motivation for this simple model. We thus use model components that would actually appear to us on the sky as unscattered point sources. More complicated source models are of course possible (for instance a source model with two components with different shapes, or having more than two components), but we will restrict ourselves to this simple two-component model to avoid overinterpretation of our measurements. Thus, the model fitting we do in this work is meant to investigate whether the non-zero closure phases we see are compatible with an observed source asymmetry in some specific direction on the sky. The possible causes of any observed asymmetry (intrinsic or scattering) will be discussed in section 5. \\

\begin{figure}
\includegraphics[width=0.48\textwidth]{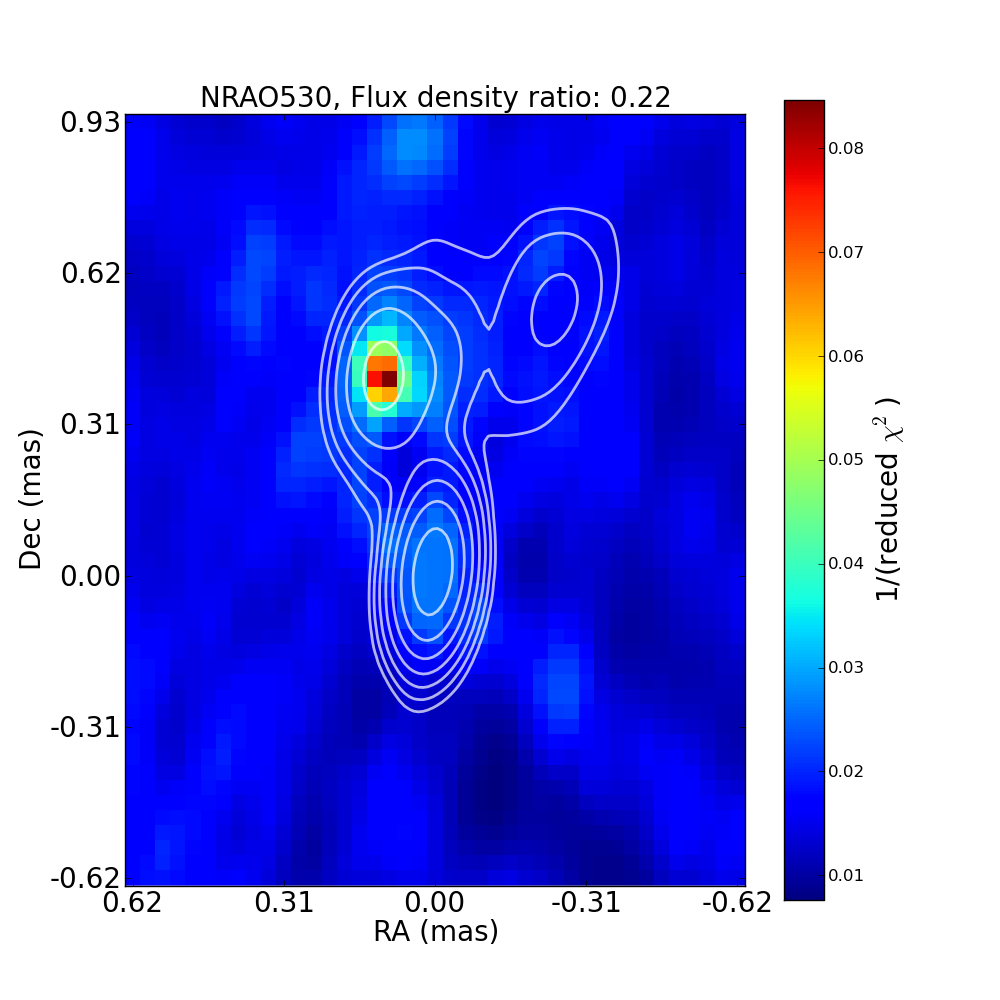}
\caption{The dependence of the reduced $\chi^2$ statistic on secondary source position for the best-fitting flux density ratio, as found for NRAO\,530 (using the closure phase measurements on all triangles). There is a strong preference for the secondary component to be situated North and slightly East of the main component (sky coordinates are expressed relative to the main component). While a 2-point-source model does not capture the structure of NRAO\,530 in detail (as reflected by the high reduced $\chi^2$ value of 11.8 for the best fit), it does capture the orientation and separation of the dominant off-center component. This is illustrated by the fact that the preferred position of the secondary component agrees with the position of the brightest off-center component found in our preliminary imaging for NRAO\,530 (overlaid contours). Contour levels are 1,2,4,8,16,32 and 64 percent of image maximum to guide the eye, with the absolute flux density calibration to be addressed in a future publication. The found flux density ratio between secondary and primary component is $\sim$0.22 for the model fit, compatible with what is found for the preliminary imaging result.}
\label{chisq-nrao530}
\end{figure}

\noindent
In the model fitting, we determine the placement on the sky and the relative flux density of a secondary source component that gives the closure phase evolution that is most consistent with our observations. To determine the best fit parameters, we use the $\chi^2$ statistic to compare the closure phases generated by the source model to the measured closure phases. In the model fitting procedure, the position of the secondary component on the sky and its flux density expressed as a fraction of the flux density of the main component are varied independently. The fitting procedure was tested using our observations of NRAO\,530, using a range of flux density ratios (0.01 to 0.99, step size 0.01) and possible secondary source component positions on the sky (up to $600\,\mu$as separation in both RA and DEC, with step size 30\,$\mu$as - forming a square grid on the sky) that was motivated by existing maps for NRAO\,530 at 3mm \citep{Lu2011}. The favoured position for the secondary source component is in excellent agreement with the source structure from our preliminary mapping results (full mapping results will be published separately), capturing the location of the dominant secondary source component. These outcomes are also in line with the previously observed structure for NRAO\,530 at 86 GHz \citep{Lu2011}, validating this simple model fitting approach. See Figure \ref{chisq-nrao530} for an illustration of this fit result.\\

\begin{figure*}
\includegraphics[width=0.48\textwidth]{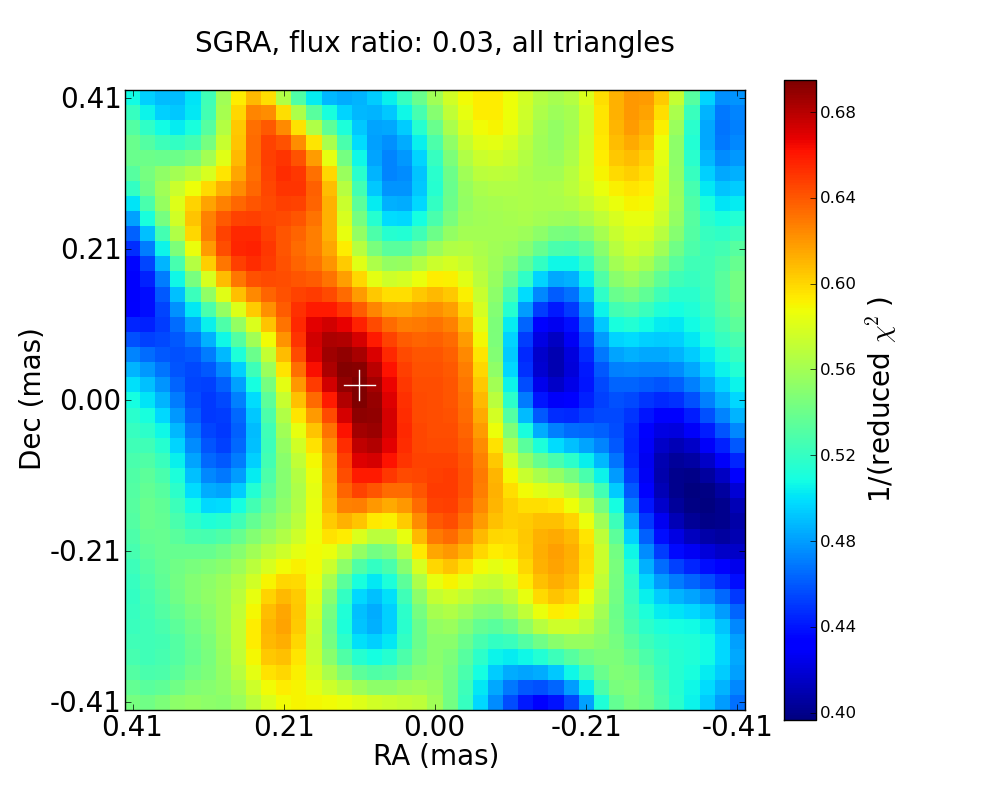}
\includegraphics[width=0.48\textwidth]{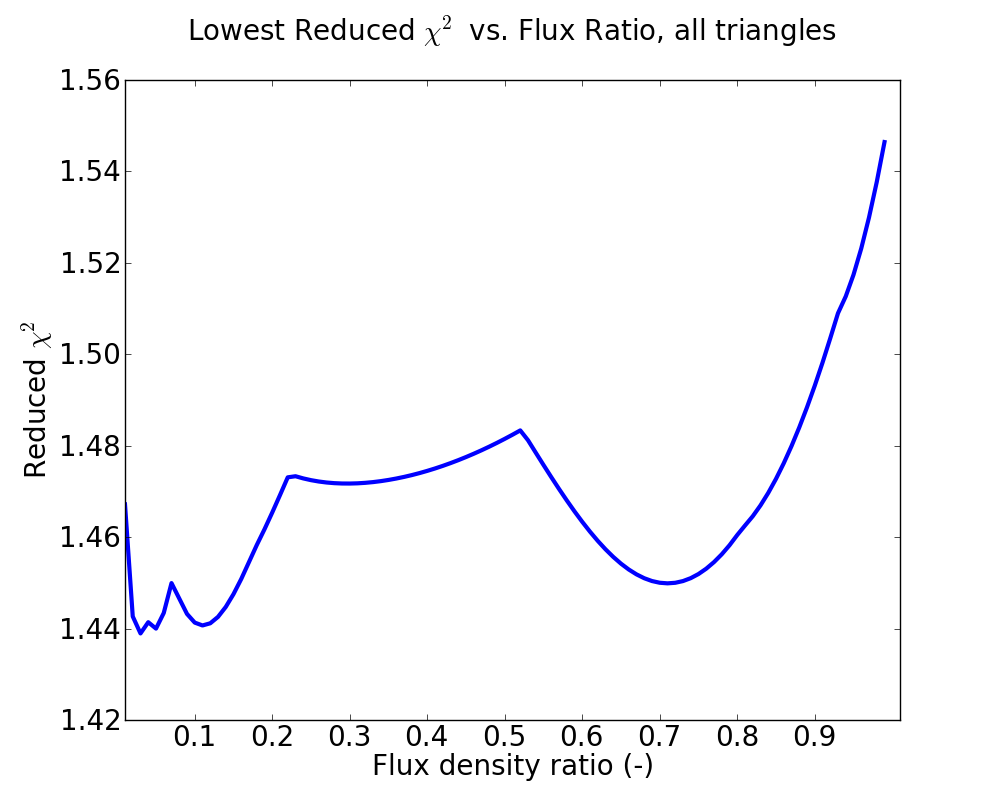}
\includegraphics[width=0.48\textwidth]{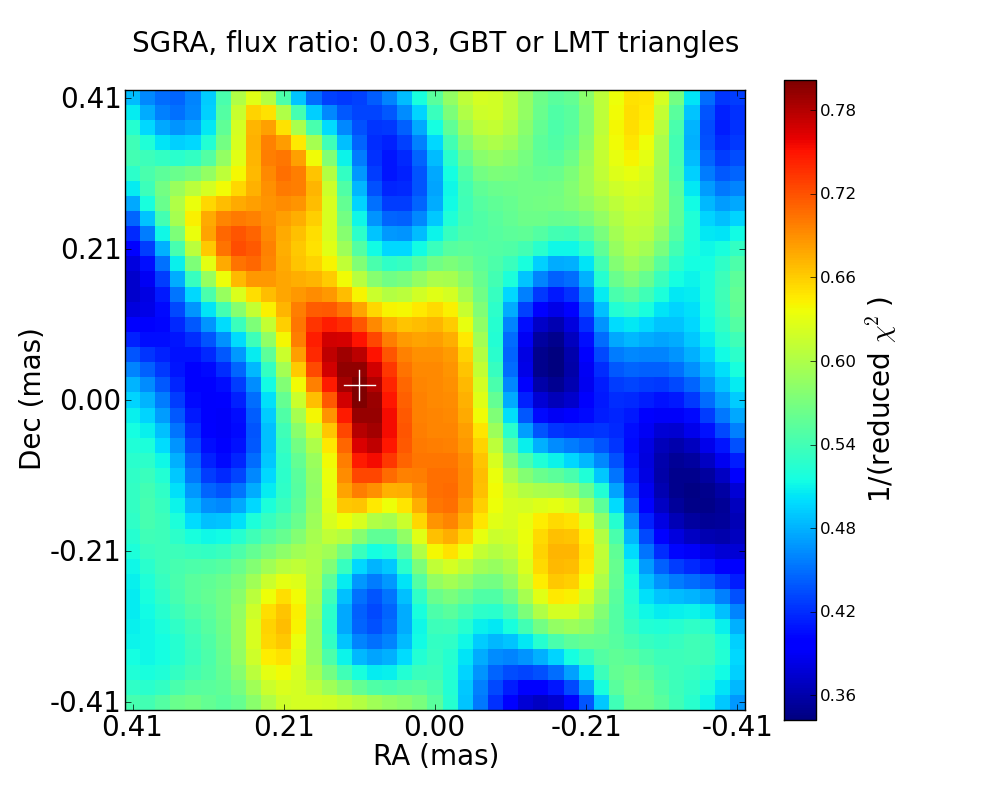}
\includegraphics[width=0.48\textwidth]{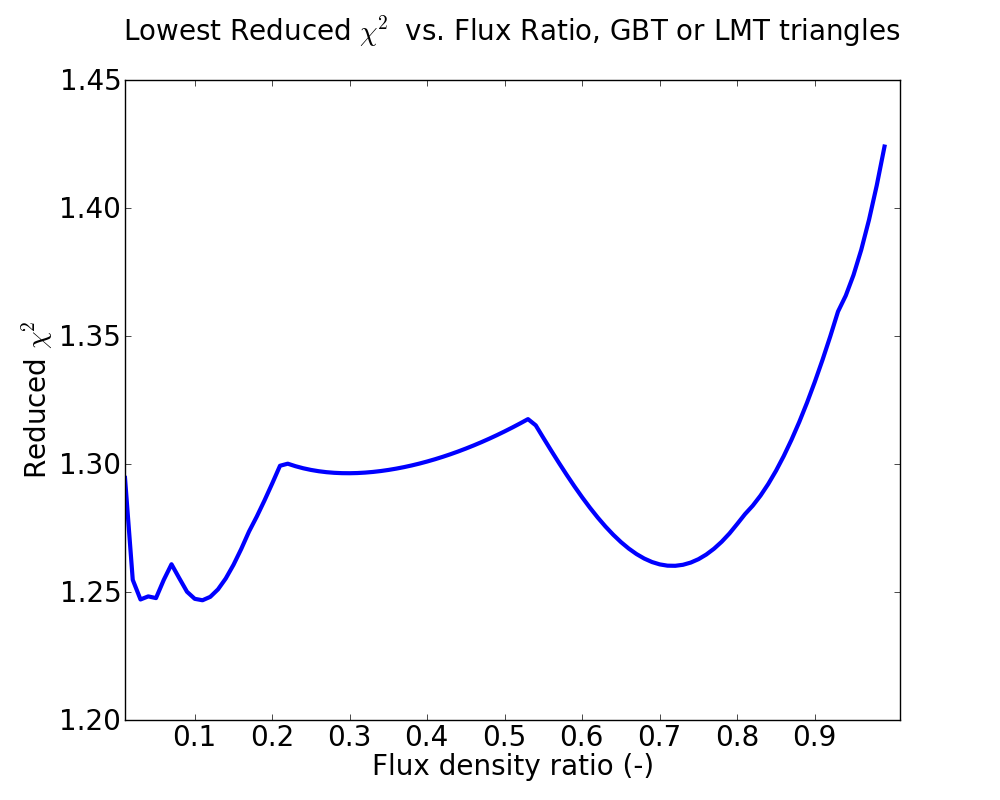}
\includegraphics[width=0.48\textwidth]{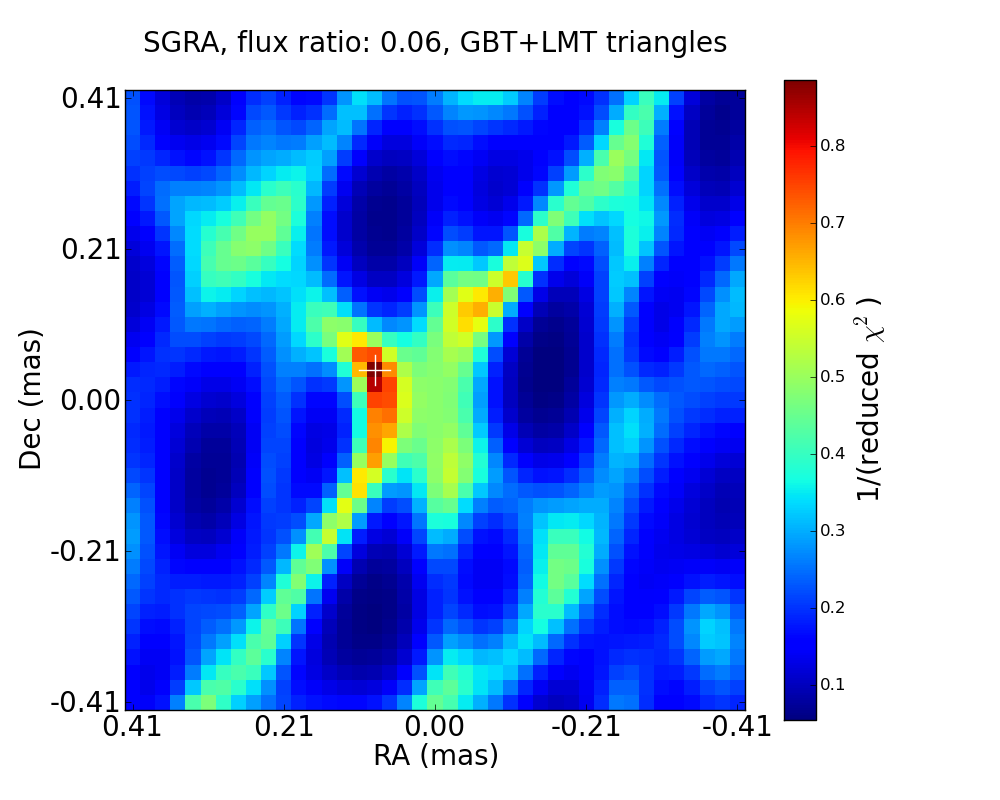}
\includegraphics[width=0.48\textwidth]{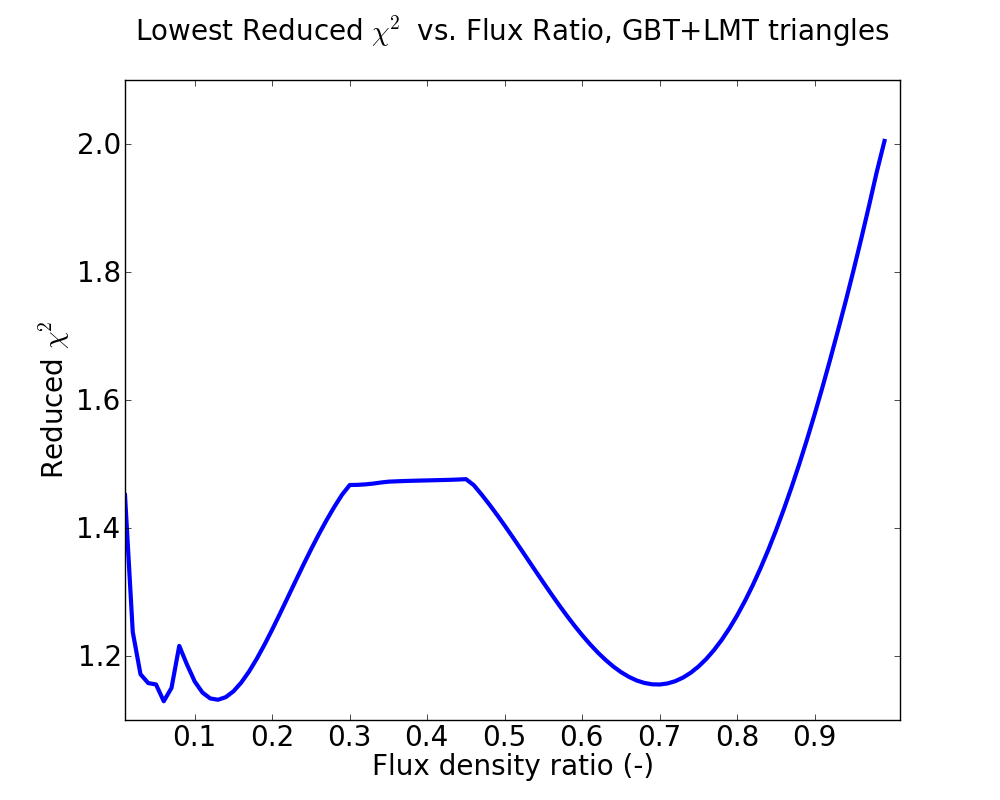}
\caption{Left column: Inverse reduced $\chi^2$ value as function of the sky position of the secondary component, for three different selections of involved triangles and shown for the flux density ratio yielding the lowest $\chi^2$ value. The position of the best fit is indicated with a white cross in each of the plots. This position is robust: for all flux density ratios, the lowest reduced $\chi^2$ score is obtained for a secondary component towards the East of the primary component. Right column: the best reduced $\chi^2$ value found for each flux density ratio. Local minima occur around flux ratios of 0.03, 0.11, and 0.70 for all triangle selections.}
\label{chisquaredplots}
\end{figure*}

\noindent
For Sgr\,A*, the position of the secondary component relative to the primary component on the sky was independently varied from $-400\,\mu$as to $400\,\mu$as with step size $20\,\mu$as in both right ascension and declination, and the dimensionless flux density ratio of the secondary to the primary component was varied from 0.01 to 0.99 in steps of 0.01. The source model used thus has three free parameters. The resulting closure phases as a function of time were simulated for all triangles and the $\chi^2$ statistic was calculated using the model curves with all of our observed data for Sgr\,A*. For practically all flux density ratios, the best-fit position on the sky for the secondary component is $\sim100\,\mu$as East of the primary component (Fig.~\ref{chisquaredplots}). The flux density ratio exhibits multiple local minima in $\chi^2$, at 0.03, 0.11 and 0.70 respectively. The flux density ratio is evidently not well-constrained by closure phases only. To constrain this flux density ratio, careful amplitude calibration of the data is needed. Results based on the fully calibrated dataset will be the subject of a separate publication. As full amplitude calibration is a tricky and involved process particularly for LMT and GBT, we have avoided relying on amplitude calibration here. While the direction of the source asymmetry on the sky is well-constrained, the uncertainty in the flux density ratio implies that there is significant uncertainty in the angular separation of the secondary component with respect to the primary.\\

\begin{figure*}
\includegraphics[width=0.48\textwidth]{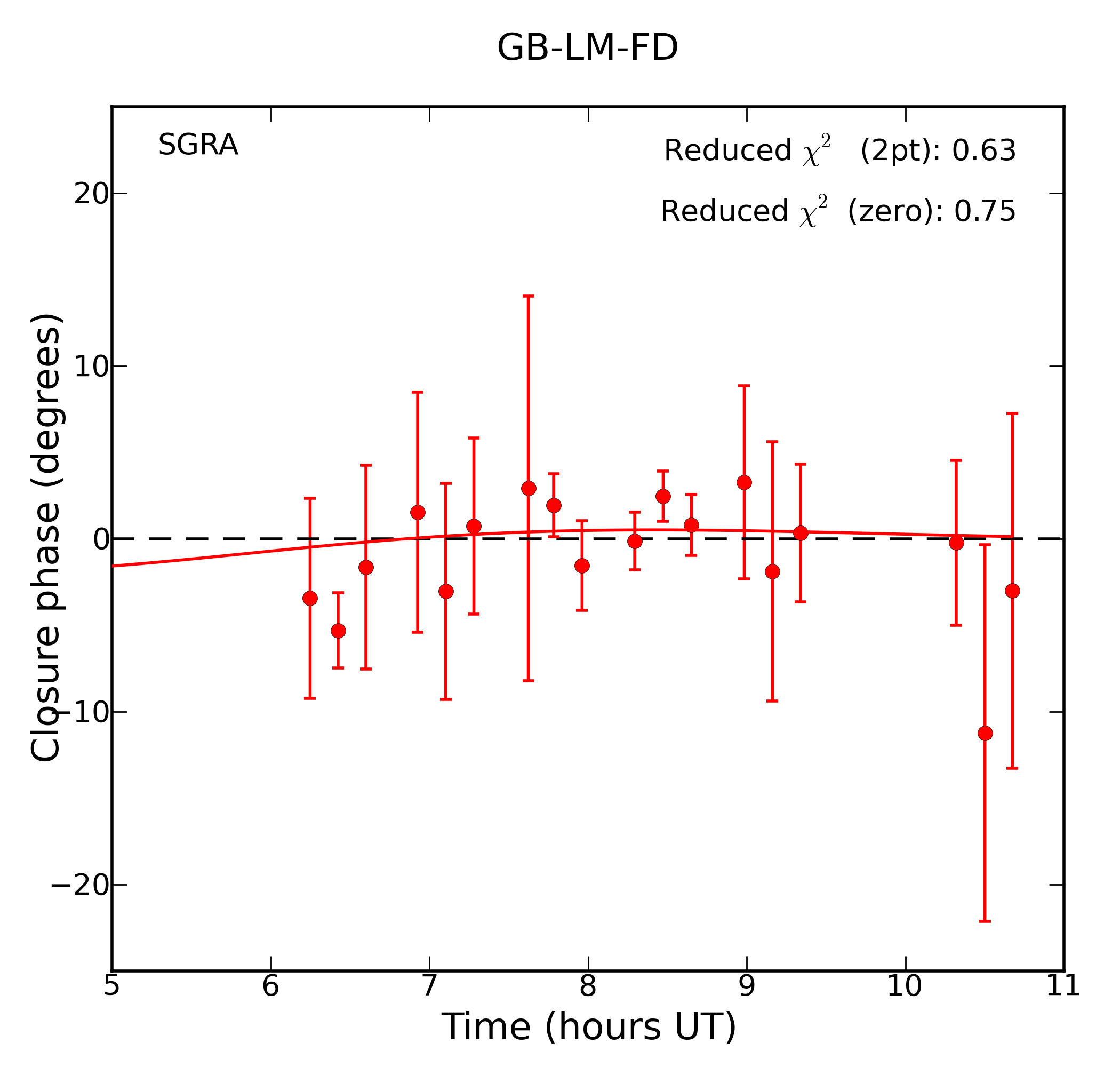}
\includegraphics[width=0.48\textwidth]{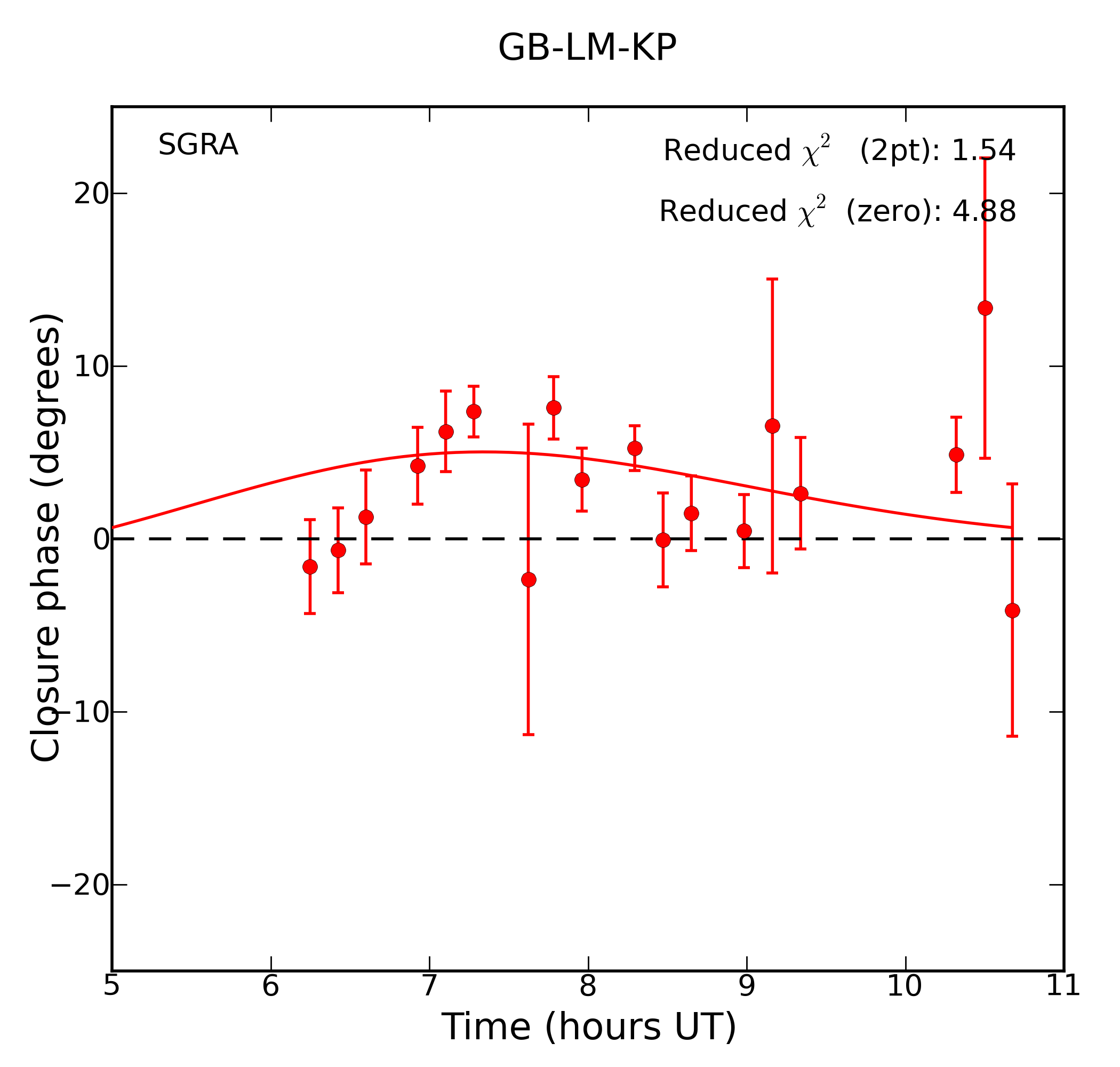}
\includegraphics[width=0.48\textwidth]{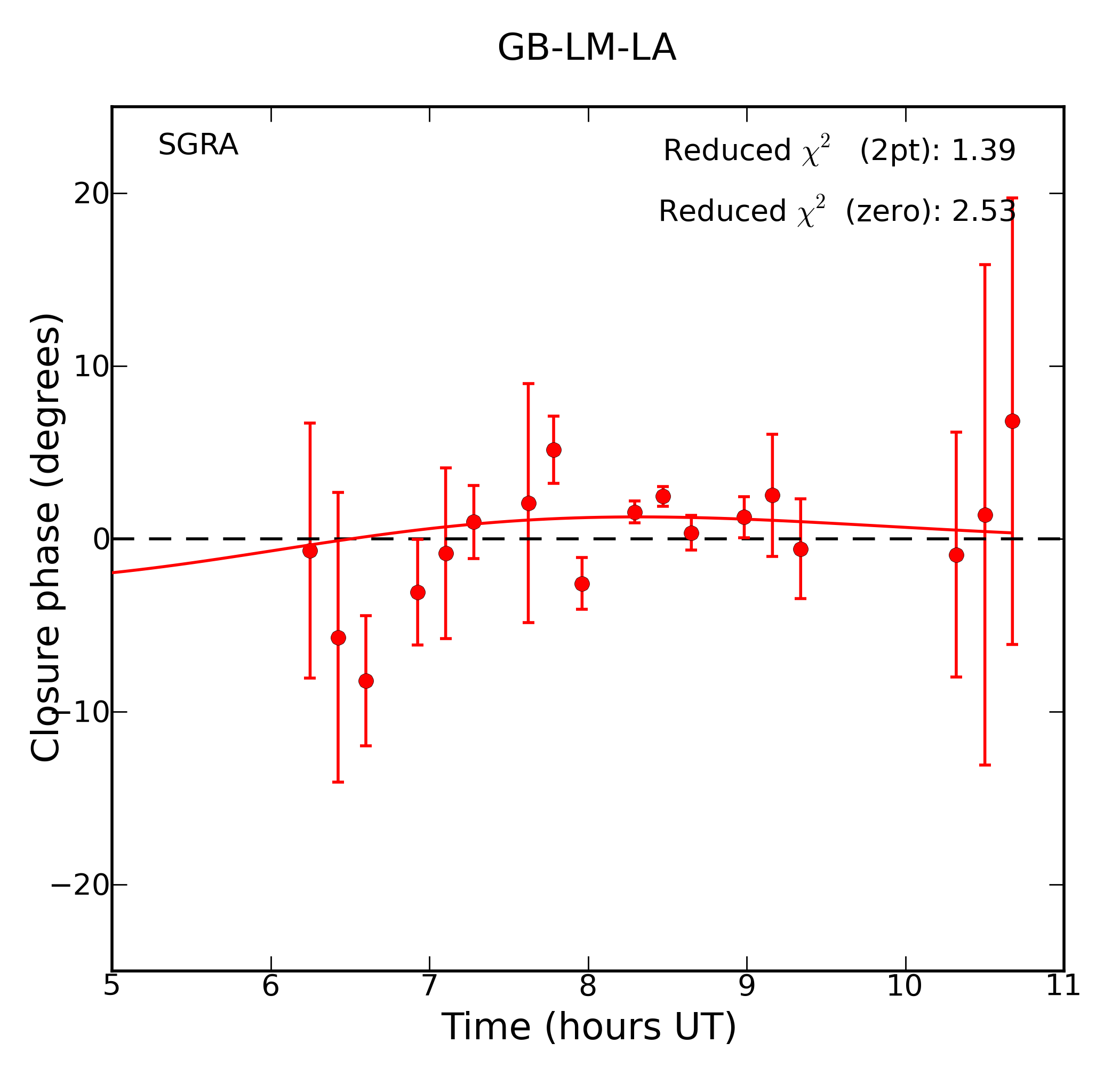}
\includegraphics[width=0.48\textwidth]{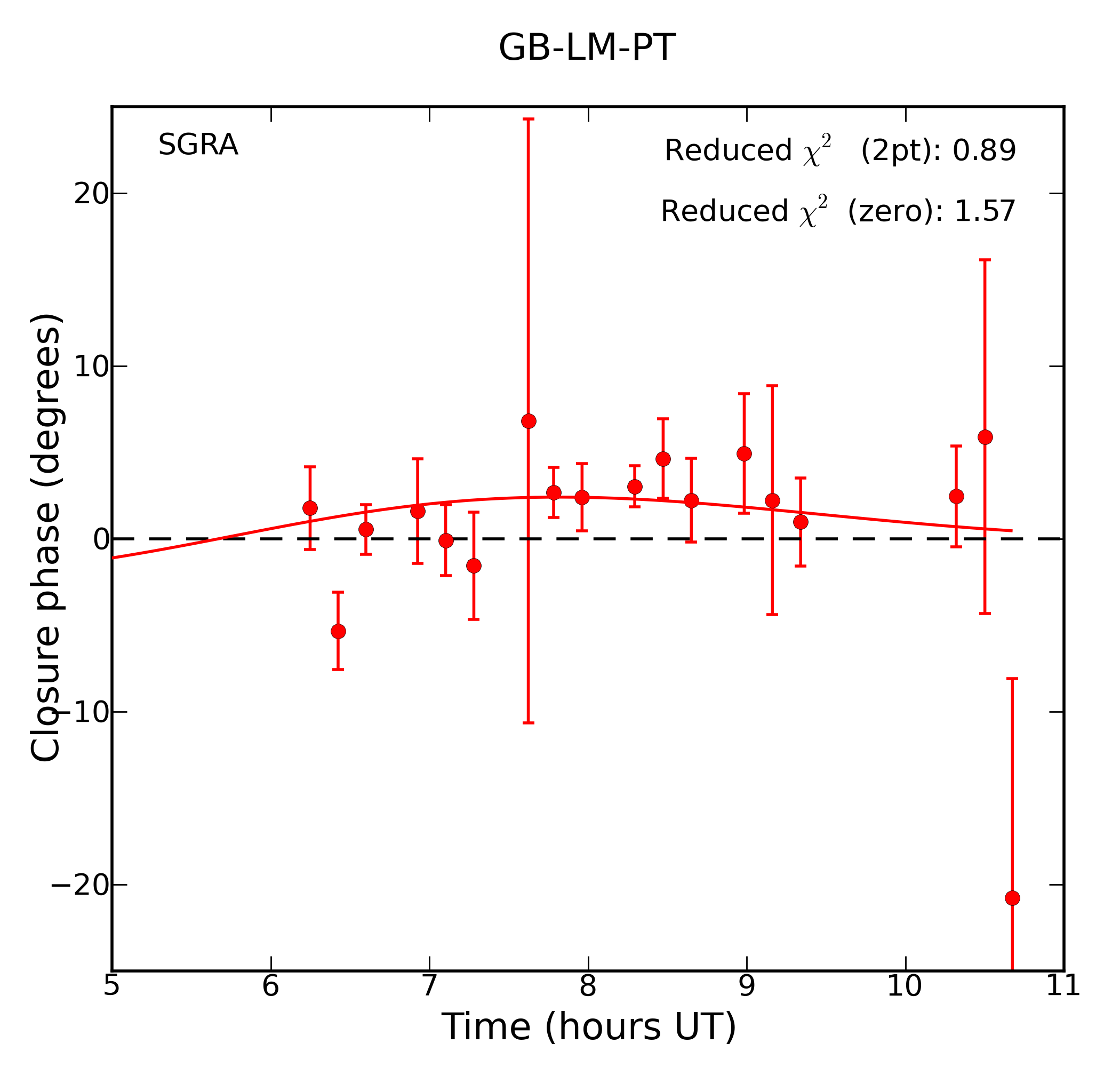}
\caption{Closure phase versus time for Sgr\,A* for the four `central' triangles in the array. All four triangles share a qualitatively similar evolution of the closure phase with time.  The two-component model closure phase curve derived from a global fit has been superimposed for each triangle. Reduced $\chi^2$ scores per triangle for both the two-component model (label `2pt') and the baseline zero closure phase model (label `zero') are indicated in the top right of every plot. The simple two-component source model matches the measurements better than the zero closure phase model in every case. The closure phase evolution for the two-component model is sensitive to the East-West extent of the triangle in a nonlinear way, hence the larger predicted closure phase deviations for the GB-KP-LM triangle.}
\label{closurephases-sgra}
\end{figure*}

\noindent
We have done this minimum $\chi^2$ search for different choices regarding the triangles included. We have considered the following options: 1) all triangles, 2) only triangles including either GBT or LMT, 3) only triangles involving both LMT and GBT, 4) all triangles without the LMT, 5) all triangles without the GBT, 6) VLBA-only triangles. We find the previously quoted secondary component position to give the lowest $\chi^2$ scores for all of these cases, with the strongest significance for case 3. It appears that inclusion of VLBA-only triangles diminishes the significance of the result, as these triangles tend to add only noise to the data to be fitted to. We show the modeled closure phase evolution for several triangles in Fig.~\ref{closurephases-sgra}, along with the reduced $\chi^2$ results for both the two-component model and the zero model. The overall $\chi^2$ scores for the best-fitting model and the zero-model can be seen in Table~\ref{table2}. The two-component model shows a better fit than the symmetric `zero' model, with the significance of this difference varying according to which set of triangles is considered. We do not expect to find reduced $\chi^2$ scores very close to 1, since the two-component model is likely an oversimplified representation of the actual source geometry. However, this simple model fit does indicate the direction on the sky for which Sgr A* shows asymmetry.\\

\begin{figure*}
\centering
\begin{minipage}{0.48\textwidth}
\includegraphics[width=\textwidth]{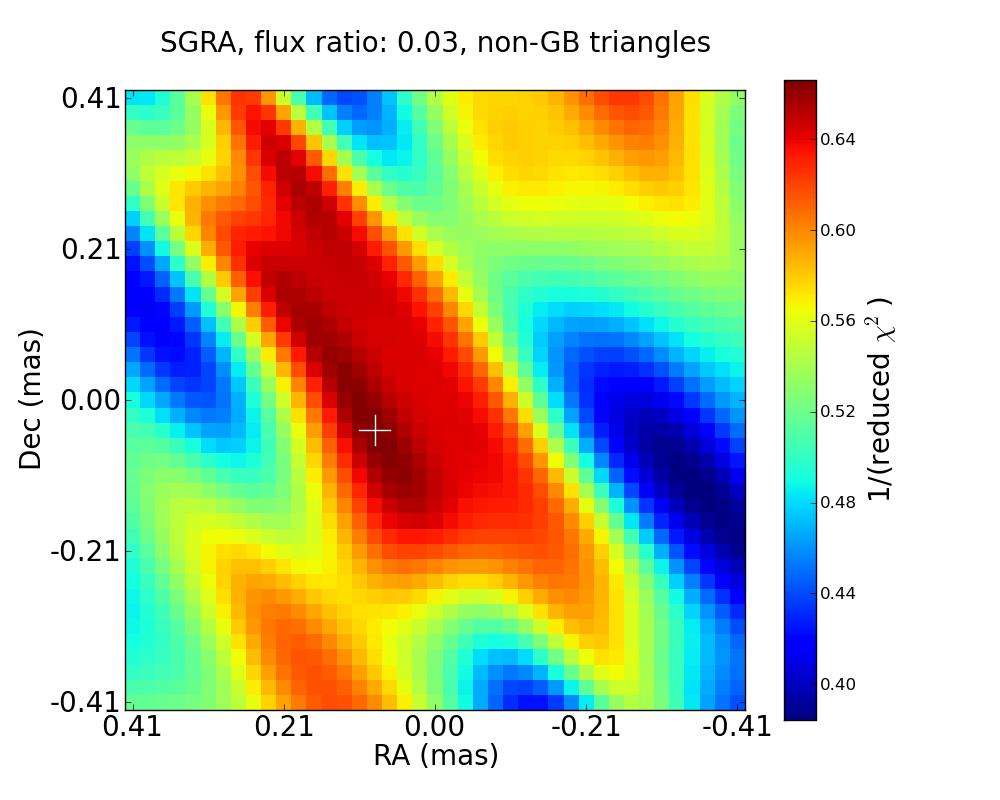}
\end{minipage}
\hfill
\begin{minipage}{0.48\textwidth}
\includegraphics[width=\textwidth]{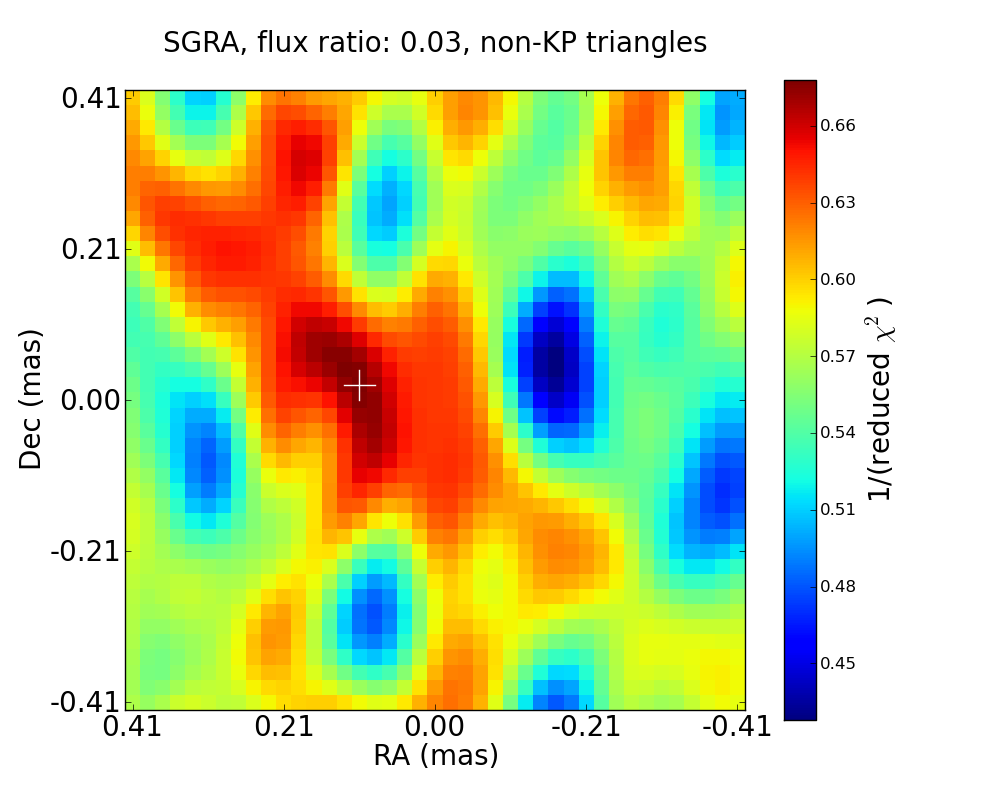}
\end{minipage}
\begin{minipage}{0.48\textwidth}
\includegraphics[width=\textwidth]{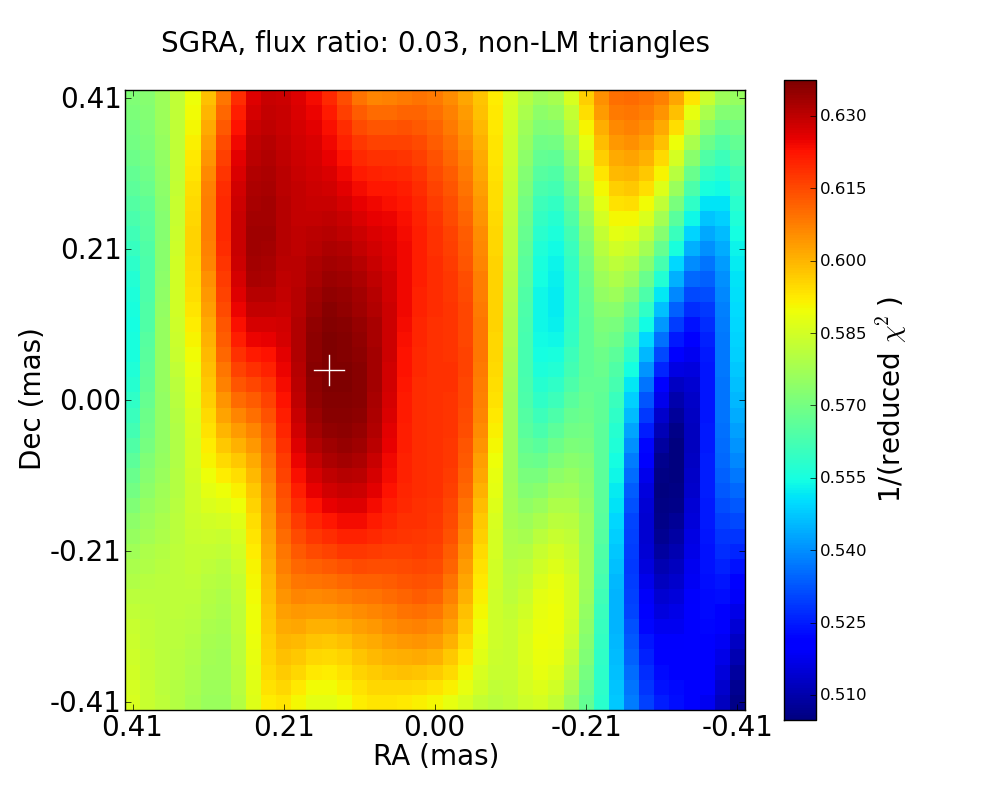}
\end{minipage}
\begin{minipage}{0.3\textwidth}
\caption{$\chi^2$ landscapes for the position of the secondary source component, considered when leaving out different key stations. Shown in the subplots are the $\chi^2$ scores when omitting all triangles with GB, KP, or LM respectively.}
\label{chisquaredplots-minus}
\end{minipage}
\hfill
\begin{minipage}{0.1\textwidth}
\end{minipage}
\end{figure*}

\noindent
The GBT-LMT-KP triangle exhibits the strongest deviations from zero in its closure phases. This gives rise to the question whether the model-fitting results are dominated by the influence of this individual triangle. To investigate this possibility, we consider the $\chi^2$ scores we get when omitting any of those 3 stations from the array, limiting ourselves to the triangles that can be formed with the other stations. The results are shown in the left column of Figure \ref{chisquaredplots-minus}. We see that the preference for an offset secondary source component to the East persists in all cases, but that the significance of the result is affected by the omission of the station in question. For instance, leaving out the LMT gives a fit result that is much less constrained in the North-South direction - as can be expected from the $(u,v)$ coverage offered by the LMT. The favoured offset position of the secondary source component also persists when any other station from the array is dropped. These results indicate that the fit results are not dominated by possible data artefacts associated with a specific station or baseline.\\

\noindent
The relatively rapid changes in the measured closure phase on the GBT-LMT-KP triangle (see Figure \ref{closurephases-sgra}) are not fully captured by the 2-component source model, and suggest a possible time-variable source structure for Sgr\,A*. Time-segmentation of the measurement data into 1-hour blocks and running the model-fit algorithm on these individual timeframes however shows no significant deviation of the secondary component in the time segment for 7 to 8 UT versus the best-fit position seen in other blocks: the found positional offsets for different time blocks are mutually compatible. This however only indicates a constant structure when the 2-point source model is assumed. More sophisticated model fits may still exhibit time-variable structure.\\

\subsection{Testing the significance of the observed asymmetry}
\begin{figure}
\includegraphics[width=0.48\textwidth]{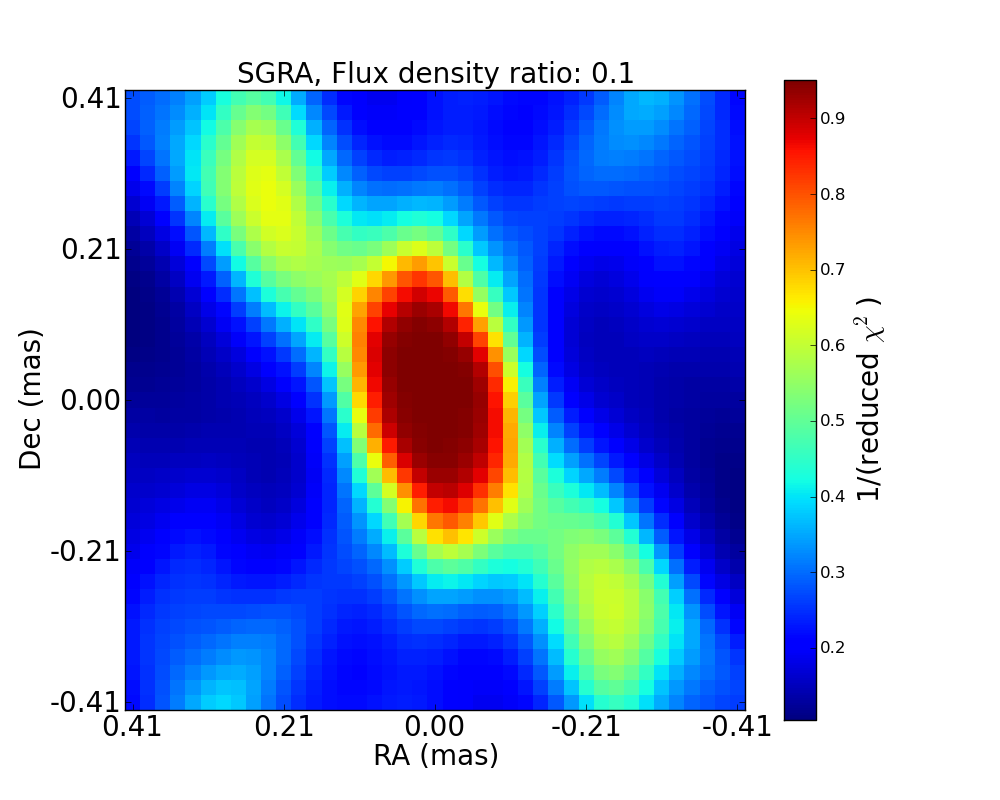}
\caption{The typical $\chi^2$ landscape for Sgr\,A* as resulting from the model fit when using the synthetic (zero-mean) closure phase data (using all triangles). Chi-squared scores do not strongly depend on flux density ratio for the synthetic data.}
\label{chisq-fake}
\end{figure}

\noindent
We need to verify that the asymmetry in Sgr\,A* as suggested by the closure phase measurements is significant. To this end, we have synthesized a control dataset in which every data point has the same measurement error as the corresponding measurement point in the original dataset. The measurement values in this control dataset have values drawn from a zero-mean normal distribution using the original measurement errors for the standard deviation. We thus get a simulated set of closure phases that corresponds to a point-symmetric source on the sky, with zero closure phases for all independent triangles to within measurement errors. Searching for the best-fitting two-component model using this simulated dataset in the way described above, we see that the best-fit $\chi^2$ is comparable to the zero-offset $\chi^2$ (see Table~\ref{table2} and Fig.~\ref{chisq-fake}). This in contrast to the results we get with the real dataset, where we see that the two-component model fit consistently shows a preference for an offset source component. For the zero-closure phase control dataset, we also see that the best $\chi^2$ value does not show a clear dependence on the flux density ratio - which is to be expected, as the best-fit position of the secondary component tends to be at the origin and hence produces zero closure phases regardless of flux density ratio.\\

\begin{figure}
\includegraphics[width=0.48\textwidth]{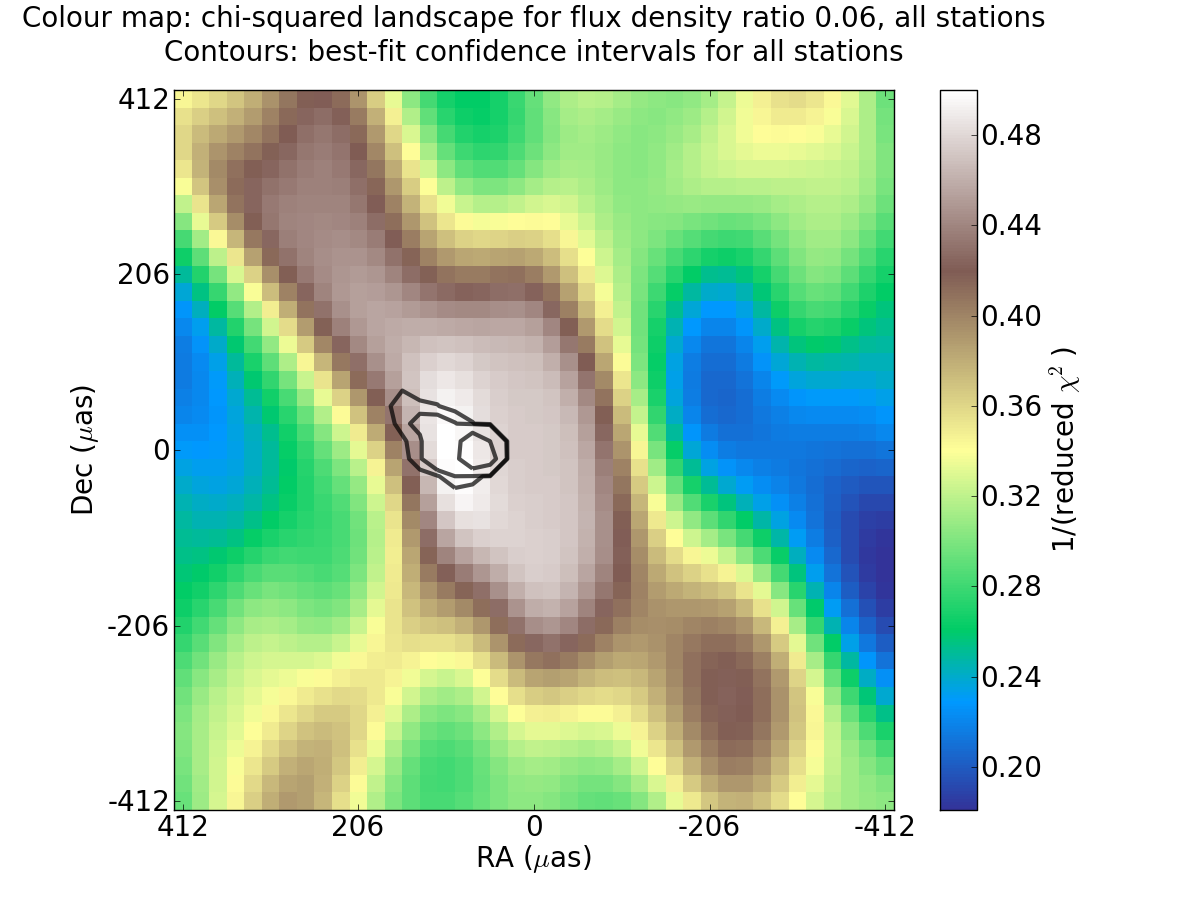}
\caption{Confidence regions (black lines) for the best-fit position of the secondary source component, obtained by bootstrapping the original closure phase dataset. The innermost contour indicates the 99\% confidence region, surrounded by the 95\% and 68\% regions respectively.}
\label{confidence-regions}
\end{figure}

\noindent
We have further assessed the uncertainty in the fitted position for the secondary source component using a bootstrapping algorithm. Bootstrapping was done by synthesizing a new closure phase dataset from the existing closure phase data by repeatedly picking measurement points at random and independently from the measured dataset and adding these to a new, synthesized dataset. The final synthesized dataset contains as many data points as the original, but typically contains multiple copies of several original measurement points and misses other original measurement points. Such a synthesized dataset was generated 1000 times and the model fitting procedure was performed on each of them. This yielded a distribution of best-fit secondary source component positions which we used to define confidence intervals on this position, see Figure \ref{confidence-regions}. The major advantage of bootstrapping is that it is robust against the presence of a subset of data points that would otherwise dominate the results of a model fitting procedure. As the result from the bootstrapping procedure agrees with the result from the original model fitting, we conclude that the asymmetry of the source we see from the original model fitting is something that is present in the dataset as a whole rather than something arising from a small selection of measurement points.\\

\begin{table*}
\caption{Sgr A* $\chi^2$ scores for the best-fit 2-component model (`2pt') and zero-closure phase model (`0'), shown for the actual closure phase measurements and for the synthetic, zero-compatible dataset. Each line in the table is valid for a different combination of selected stations in the array. Columns headed $\chi^2/\textrm{d.o.f.}$ indicate the $\chi^2$ value and the degrees of freedom, while columns headed $\chi^2_{\textrm{red}}$ give the reduced $\chi^2$ figures for convenience.}
\label{table2}
\begin{tabular}{l|llll|llll}
Stations in triangles & \multicolumn{4}{|c}{Measurements} & \multicolumn{4}{|c}{Synthetic data} \\
\, & $\chi^2$/d.o.f. (2pt) & $\chi_\mathrm{red}^2$ (2pt) & $\chi^2$/d.o.f. (0)  & $\chi_\mathrm{red}^2$ (0) & $\chi^2$/d.o.f. (2pt) & $\chi_\mathrm{red}^2$ (2pt) & $\chi^2$/d.o.f. (0)  & $\chi_\mathrm{red}^2$ (0) \\
\hline
All                        &2252/1564 & 1.440 & 2432/1567 & 1.552 & 1576/1564 & 1.008 & 1580/1567 & 1.009 \\
GBT and/or LMT   & 1116/889 & 1.255 & 1283/892 & 1.438 & 886/889 & 0.996 & 889/892 & 0.997 \\
both LMT and GBT & 135/118 & 1.140 & 241/121 & 1.994 & 104/118 & 0.884 & 109/121 & 0.904 \\
no LMT                & 1608/1025 & 1.569 & 1657/1028 & 1.612 & 1044/1025 & 1.019 & 1050/1028 & 1.021 \\
no GBT                 & 1624/1088 & 1.493 & 1683/1091 & 1.543 & 1110/1088 & 1.020 & 1113/1091 &1.020 \\
no KP                   & 1499/1025 & 1.463 & 1590/1028 & 1.547 & 1027/1025 & 1.002 & 1030/1028 & 1.002 \\
VLBA only            & 1120/671.0 & 1.669 & 1149/674.0 & 1.705 & 688/671.0 & 1.026 & 691/674.0 & 1.026 \\

\end{tabular}
\end{table*}

\section{Discussion}

We argue that VLBI observations at 3\,mm probe a sweet spot in frequency, making them ideally suited to investigate the source structure and size. This is on one hand because the influence of interstellar scattering diminishes strongly with increasing frequency - observations at lower frequencies are more strongly influenced by scattering effects (leaving little to no opportunity to study intrinsic source structure). On the other hand, observations at higher frequencies are expected to show a source geometry that is increasingly dominated by strong lensing effects around the black hole shadow. Both of these cases throw up obstacles when studying the geometry of the inner accretion flow itself. Observations at 3\,mm thus mitigate some of the complexities of interpretation associated with observations at longer and shorter wavelengths: while the effects of interstellar scattering still cannot be ignored at 3mm, intrinsic source geometry can be distinguished from scatter-induced features given multiple observations.\\

\noindent
We deduce that Sgr\,A* exhibits asymmetry in the East-West direction, with a source geometry that features a weaker source component about $100\,\mu\textrm{as}$ to the East (PA: $\sim$90$^{\circ}$) of the main source (where we note that the separation is poorly constrained). Earlier observations at 86\,GHz than those done over the last year were limited by the available $(u,v)$ coverage, and thus the best intrinsic source sky models were limited to anisotropic, but symmetrical (2D) Gaussians. The scattering kernels were modeled as Gaussians as well, allowing subtraction in quadrature of the scattering kernel from the best-fit observed source Gaussian. This approach has yielded an intrinsic source size that showed an elongated source shape along an approximately East-West direction. We note that the best-fit position for the secondary component falls along the major axis of the scattering ellipse as it was measured by \citet{Bower2014, Bower2015} and is also compatible with the previously observed intrinsic elongation of the source quoted in these publications.\\

\noindent
These observations cover a single epoch and were done in a single frequency band and in a single polarisation (LCP), which complicates interpretation of the observed asymmetry. On one hand, interstellar scattering of the source image can introduce small-scale scintels whose ensemble average influences the observed brightness distribution \citep{Gwinn2014, Johnson2015-2} and that may be responsible for the occurrence of non-zero closure phases \citep{Ortiz2016}. The time scale for the scattering geometry to change significantly ($\sim$weeks) is thought to be much longer than the length of one observation ($\sim$hours), causing the source image to be affected by an effectively static scattering screen that may induce asymmetry in the observed image. On the other hand, the observed asymmetry may be intrinsic to the source itself. Observations at different frequencies (e.g., at 230\,GHz and 43\,GHz) and performed at different epochs (separated in time by months) are therefore crucial in interpreting the character of this observed asymmetry.\\

\noindent
The 86\,GHz observations published by \citet{Ortiz2016} do show non-zero closure phases, but these have been interpreted consistently as arising from interstellar scattering effects. As such, no dedicated closure-phase modelling comparable to the analysis presented in this work was performed. Those data are separated in time from the observation we report in this work by approximately one month (April 27th vs May 23rd, 2015). Future studies of the non-zero closure phase evolution with time will help to distinguish its origin: if the observed asymmetry is persistent across both datasets, the case for an intrinsic cause of the asymmetry will be bolstered as scattering effects are expected to vary over shorter timescales \citep{Johnson2015-2}. Conversely, if the earlier data show a different asymmetry from what we find here the likely cause for it will be confirmed as being interstellar scattering.\\

\noindent
Interestingly, an East-West asymmetry in Sgr\,A* is also suggested by closure phase results from measurements taken with the Event Horizon Telescope at 230\,GHz, in the Spring of 2013 \citep{Fish2016}. The observations presented in that work show closure phases at 1.3\,mm that are comparable in magnitude to the values we have measured at 3.5\,mm, suggesting a similar degree of source asymmetry in both observed emission patterns. While a source model with disconnected components is not necessarily favoured by the EHT data, fit results using a model consisting of 2 point sources suggest a preference for an East-West asymmetry in that dataset. It is somewhat surprising that the persistent asymmetry at 230\,GHz is oriented along the same direction on the sky as the asymmetry found in this work. At 230\,GHz a persistent asymmetry in the source image is expected, and is thought to be caused by the Doppler boosting of emission from one side of the inner accretion flow with a velocity component along our line of sight \citep{Broderick2016}. Conversely, at 86\,GHz this effect is not expected to be a dominant contribution to source asymmetry - rather, the main part of any intrinsic asymmetry is expected to be a consequence of the relative brightness of the inner accretion flow versus emission from the footpoints of a compact jet component \citep{Moscibrodzka2014}. In the context of this model, the similar orientation of the asymmetry in the 230\,GHz and 86\,GHz observations cannot be reconciled if both are assumed to be intrinsic to the source.\\

\noindent
For spectrally fitted jet models, a significant component of the emission at 86\,GHz is generated around the jet base \citep{Moscibrodzka2013}, causing the corresponding source image to exhibit an asymmetry that is aligned with the jet axis to within $\sim$20 degrees. In this context the results from this work, when combined with other existing measurements of Sgr\,A* closure phases at 3\,mm, offer an appropriate starting point for a more extended model fitting procedure, where the raytraced results from GRMHD simulations can be compared to the constraints on the observed source geometry. An analogous analysis has been performed on the published 230\,GHz closure phase measurements in \citep{Broderick2016}, where the measurements have been interpreted within the context of a particular theoretical source model. This more elaborate model fitting procedure using the full available body of 86\,GHz closure phase data is the focus of a separate publication that currently is in preparation.\\

\section{Summary and conclusions}

We have performed an observation of Sgr\,A* at 86\,GHz, using the VLBA, the GBT, and the LMT. Elementary model fitting of a multicomponent source geometry to the closure phases from this dataset shows a preference for an Eastern secondary source component at an on-sky separation of $\sim$100\,$\mu$as from the primary component. This asymmetry, when considered as a standalone observation, may be explained by interstellar scattering effects. However, this does not exclude the possibility of the observed asymmetry being intrinsic to the source.\\

\noindent
The results by \citet{Fish2016} at 230GHz , \citet{Ortiz2016} at 86GHz, and \citet{Rauch2016} at 43 GHz indicate asymmetric emission of Sgr A* at different frequencies and over different time periods. In particular the closure-phase measurements performed at 230 and 43 GHz point towards a similar East-West asymmetry as was found in the dataset presented in this work. The similar orientation of this asymmetry across these different wavelengths is a puzzling result, and future analysis of 86\,GHz VLBI measurements done at different times will help to pin down the origin of these observed non-zero closure phases.

\section*{Acknowledgements}
We wish to express our gratitude to the MIT Haystack team (Lindy Blackburn, Laura Vertatschitsch, Jason Soohoo), who installed the recording system at LMT and who have played an instrumental role in making VLBI measurements possible at LMT. We thank Frank Ghigo at GBT for his help in obtaining the system temperature measurements. We thank Michael Johnson for illuminating discussions on the scattering screen and on closure phase statistics, and we appreciate the input on the draft we received from Eduardo R\'os. This work is supported by the ERC Synergy Grant “BlackHoleCam: Imaging the Event Horizon of Black Holes” (Grant 610058). A.H., L.L., and G.N.O.-L. acknowledge the financial support of CONACyT, Mexico and DGAPA, UNAM. 

\bibliographystyle{apalike}
\bibliography{mnemonic,aaabbrv,closure}

\bsp

\clearpage

\clearpage

\clearpage

\clearpage

\clearpage

\clearpage

\clearpage

\clearpage

\label{lastpage}

\end{document}